\pdfoutput=1
\documentclass[iop]{emulateapj}
\usepackage{hyperref}

\usepackage{booktabs}
\usepackage[usenames,dvipsnames]{color}
\usepackage{rotating}
\usepackage{epstopdf}
\usepackage{color}
\usepackage{multirow}
\usepackage[para,online,flushleft]{threeparttable}

\newcommand{\halpha}{H$\alpha$}
\newcommand{\hbeta}{H$\beta$}
\newcommand{\lya}{\ensuremath{\mathrm{Ly}\alpha}}
\newcommand{\lsim}{\raise0.3ex\hbox{$<$}\kern-0.75em{\lower0.65ex\hbox{$\sim$}}}

\newcommand{\msun}{M$_{\odot}$}
\newcommand{\pom}{\,$\pm$\,}

\newcommand{\kms}{km\ s$^{-1}$}

\newcommand{\Ha}{H$\alpha$}

\newcommand{\HI}{\mbox{H\thinspace\footnotesize I}}
\newcommand{\HII}{\mbox{H\thinspace\footnotesize II}}

\usepackage{fancyheadings}
\usepackage{amsmath}
\usepackage{amssymb}
\usepackage{subfigure}
\usepackage{graphicx}
\usepackage{comment}
\usepackage{natbib}
\bibpunct{(}{)}{;}{a}{}{,}
\usepackage{appendix}

\begin{document}


\title{Detection of HI in Emission in the Lyman Alpha Emitting Galaxy Haro 11}

\author{Stephen A. Pardy} 
\affil{Department of Astronomy, University of Wisconsin-Madison, 475 North Charter
  Street, Madison, WI 53706, USA} 
\email{spardy@astro.wisc.edu}
\author{John M. Cannon}
\affil{Department of Physics \& Astronomy, Macalester College, 1600 Grand Avenue, 
Saint Paul, MN 55105}
\author{G{\"o}ran {\"O}stlin}
\affil{Department of Astronomy, Oskar Klein Centre, Stockholm University, 
AlbaNova University Centre, SE-106 91 Stockholm, Sweden}
\author{Matthew Hayes}
\affil{Department of Astronomy, Oskar Klein Centre, Stockholm University, 
AlbaNova University Centre, SE-106 91 Stockholm, Sweden}
\author{Nils Bergvall }
\affil{Department of Physics and Astronomy, Uppsala University, Box 515, 751 20 Uppsala, Sweden}

\begin{abstract}

We present the first robust detection of \HI\ 21 cm emission in the blue compact galaxy Haro 11 using the 100m Robert C. Byrd Green Bank Telescope (GBT). Haro 11 is a luminous blue compact galaxy with emission in both Lyman Alpha and the Lyman continuum. We detect (5.1 \pom\ 0.7 $\times$10$^8$) \msun\ of \HI\ gas at an assumed distance of 88 Mpc, making this galaxy \HI\ deficient compared to other local galaxies with similar optical properties. Given this small \HI\ mass, Haro 11 has an elevated M$_{H2}$/M$_{HI}$ ratio and a very low gas fraction compared to most local galaxies, and contains twice as much mass in ionized hydrogen as in neutral hydrogen. The \HI\ emission has a linewidth of 71 \kms\ and is offset 60 \kms\ redward of the optical line center. It is undergoing a starburst after a recent merger which has elevated the star formation rate, and will deplete the gas supply in $<$ 0.2 Gyr. Although this starburst has elevated the SFR compared to galaxies with similar \HI\ masses and linewidths, Haro 11 matches a trend of lower gas fractions toward higher star formation rates and is below the general trend of increasing \HI\ mass with increasing luminosity. Taken together, our results paint Haro 11 as a standard low-mass galaxy that is undergoing an unusually efficient star formation episode.

\end{abstract}

\section{Introduction}
\label{sec:intro}

Haro 11 is a luminous blue compact galaxy (LBCG), a class of very bright and blue galaxies with intense star formation. Rare in the local universe \citep{Werk:2004de}, these LBCGs are typically gas rich \citep{Garland:2004ec} and likely form from merger events \citep{Ostlin:2001ch, Bekki:2008jt}. Haro 11 is a powerful emitter across the electromagnetic spectrum. It is classified as a Luminous InfraRed Galaxy (LIRG) with an infrared luminosity of $>$ 10$^{11}$L$_{\odot}$, and shows signs of strong star formation likely due to a recent merger \citep{Bergvall:2000tg, Ostlin:1999cl, Ostlin:2015tc}. Haro 11 also shows emission in \lya\ from some of its star forming regions \citep{Hayes:2007hk} and is one of a few currently known Lyman continuum emitting galaxies in the local universe \citep{Bergvall:2006ib, Leitet:2013cw}.

The detection of \lya\ in emission in Haro 11 is particularly interesting. \lya\ is an energetic star formation tracer that, in theory, can probe the stellar processing in the early epochs of the universe. In practice, this line suffers from strong resonance scattering, uncertain extinction properties, and the fact that it is unobservable from the ground due to atmospheric absorption. 

The scattering and extinction process is further complicated by a number of confounding factors. Dust content and properties alone cannot explain the observed extinction of \lya\ (e.g. \citealt{1996ApJ...466..831G, Atek:2009hra}); instead, trends are apparent between the neutral gas content and geometry \citep{2004ApJ...608..768C, 1998A&A...334...11K}. Super bubbles formed after star formation episodes can clear a path for escaping \lya\ photons by shifting \HI\ atoms into a different rest frame \citep{1998astro.ph..9096K, TenorioTagle:1999bx}. \lya\ photons can also scatter while avoiding dust grains and spread into large halos \citep{2013ApJ...765L..27H, Hayes:2015wy}.

Many of these hypotheses are now being tested by targeted observations using the Hubble Space Telescope \citep[HST;][]{Wofford:2013hg, Ostlin:2014vk}. In particular, the Lyman Alpha Reference Sample \citep[LARS;][]{Ostlin:2014vk, Hayes:2014jv, Pardy:2014ir, Guaita:2015kr, RiveraThorsen:2015ct} targeted 14 low-z potential \lya\ emitters based on their ultraviolet luminosities and \halpha\ equivalent widths. Initial observations of the LARS galaxies with the HST were followed up with multi-wavelength observations of the gas and dust content, including neutral hydrogen in absorption \citep{RiveraThorsen:2015ct} and emission \citep{Pardy:2014ir}. Combining observations of \lya\ and \HI\ emissions allow for direct comparisons and testing of the scattering effects at work. When combined with a dust tracer like \halpha/\hbeta, this technique can also probe the extinction effects at work \citep{Hayes:2014jv}.

\begin{figure}[h] 
   \centering
   \includegraphics[width=3.5in]{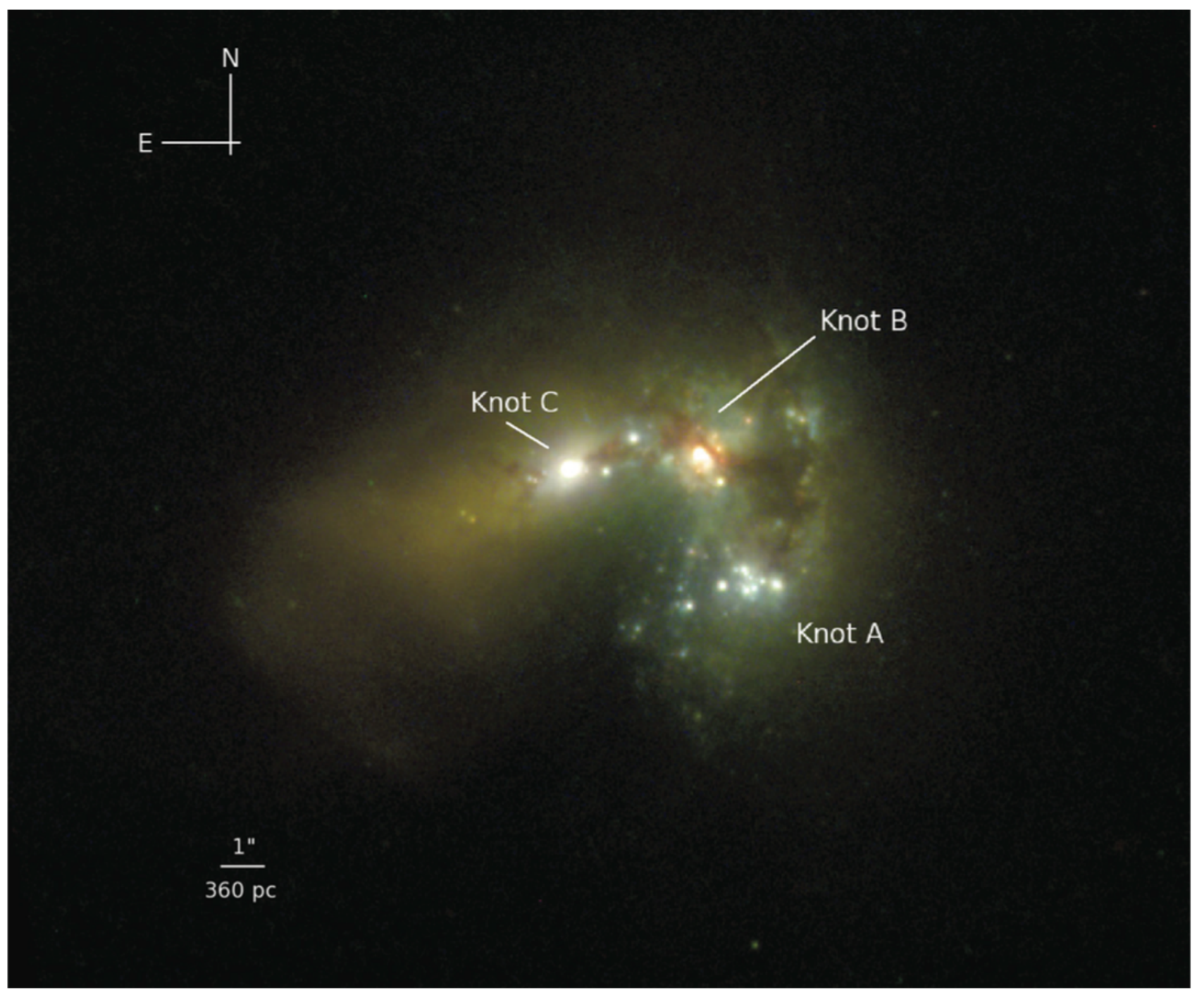} 
   \caption{HST image of Haro 11 and its three primary star-forming knots. Figure from \citet{Adamo:2010jv}. Knot C is the only knot with \lya\ emission, even though it shares very similar dust content with knot B  \citep{Atek:2008hn}, which also has a stronger outflow of interstellar gas \citep{Sandberg:2013hw}.}
   \label{fig:Haro11Img}
\end{figure}

Given its proximity and star formation intensity, Haro 11 is a prime candidate for studies of \lya\ radiative transfer. The galaxy has three primary star forming knots A, B, and C \citep[see \autoref{fig:Haro11Img};][]{Vader:1993fk, 2003ApJ...597..263K, Adamo:2010jv}. One of these knots, knot C, shows \lya\  emission, while the other two show absorption. This is striking because knots C and B have very similar dust content and extinction values \citep{Atek:2008hn}. Further complicating matters, knot B has a stronger outflow of interstellar gas, yet no sign of \lya\ emission \citep{Sandberg:2013hw}.

Haro 11's ISM properties have been extensively studied. \citet{Cormier:2014il} studied the molecular gas content of Haro 11 using both CO gas and IR dust emission. That work found a total H$_2$ mass of between 2.5$\times$10$^8$ and 3.6$\times$10$^9$ depending on the tracer used. \citet{James:2013ia} used a variety of metallicity tracers to find a range of metallicities across the star forming knots from 12 + log(O/H) = 8.25 \pom\ 0.15 and Z/Z$_{\odot}$ = 0.35 in Knot B, down to 8.09 \pom\ 0.23 (Z/Z$_{\odot}$ = 0.24) in Knot A, and 7.80 \pom\ 0.13 (Z/Z$_{\odot}$ = 0.12) in Knot C. 

Yet, to date there has been no direct detection of 21 cm emission for Haro 11. \citet{Bergvall:2000tg} first placed an upper limit on \HI\ emission of $\le$ 10$^8$ \msun. \HI\ was then seen in absorption by \citet{MacHattie:2014ipa}, giving a mass range of (3-10)$\times 10^8$ \msun, with an upper limit on the emission of M$_{HI} \le 1.7 \times 10^9$\msun. This absorption measurement assumed a spin temperature between 91 and 200 K in the optically thin regime. In this paper we present the first robust detection of the \HI\ spectral line in emission. In \autoref{sec:obs} we discuss the observations and data reduction. In \autoref{sec:results} and \autoref{sec:discussion} we present the results and interpret them in the context of LARS and other \lya\ emitters.

Throughout this paper we assume a value of H$_0$ = 70.2 \pom\ 1.4 \kms\ Mpc$^{-1}$ \citep{2011ApJS..192...18K}.

\section{Observations}
\label{sec:obs}

We observed Haro 11 with the National Radio Astronomy Observatory 100m
Robert C. Byrd Green Bank Telescope (GBT\footnote{The National Radio
  Astronomy Observatory is a facility of the National Science
  Foundation operated under cooperative agreement by Associated
  Universities, Inc.}) in four 2.5 hour sessions under project 14B-306 (P.I. Pardy) with the VEGAS (Versatile GBT Astronomical Spectrometer) backend. The native resolution of the spectrum was 0.1 \kms. We obtained both XX and YY polarizations, which we averaged together for the final data.
  
  \begin{figure}[h] 
   \centering
   \includegraphics[width=3.5in]{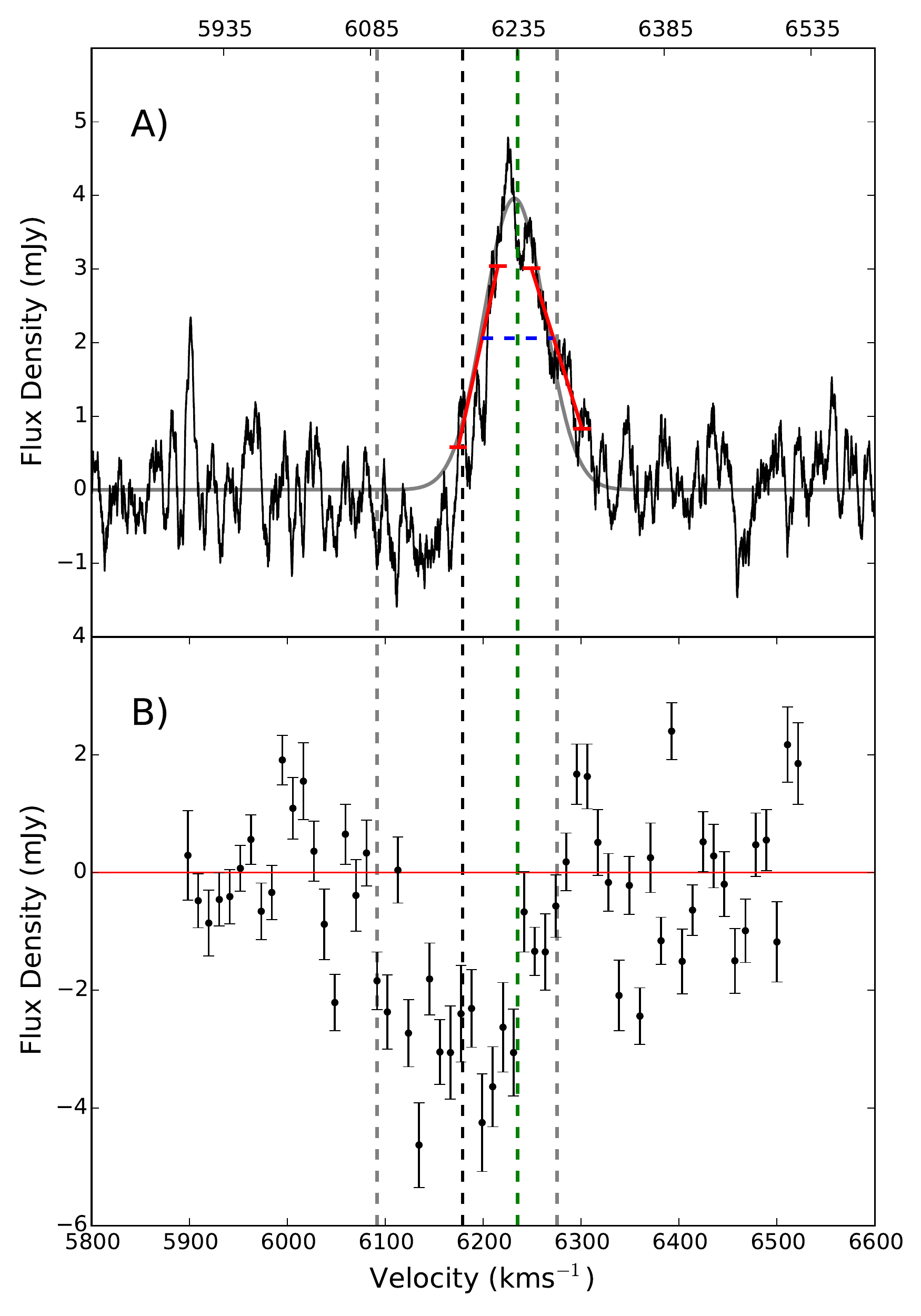} 
   \caption{\textbf{A)} Single dish \HI\ spectrum from the GBT. The spectrum is centered on the central velocity of 6236\kms\ (vertical green dash line). The red solid lines show the fits to the sides of the spectrum used to measure the velocity width at 50\% of maximum (horizontal blue dot-dash line). The solid gray line shows a single gaussian fit which is used to guide the eye and defines the velocity region to integrate over. The black and gray vertical dotted lines show the velocity center and total width of the absorption spectrum from \citet{MacHattie:2014ipa}. \textbf{B)} \HI\ absorption spectrum from \citet{MacHattie:2014ipa} (reproduced with permission from their figure 2). The solid black line shows their measured velocity of 6179 \kms.}
   \label{fig:spectrum}
\end{figure}

All reductions were performed in the IDL environment\footnote{Exelis
  Visual Information Solutions, Boulder, Colorado}, using the GBTIDL
package designed at NRAO following the reduction strategy of \citet{Pardy:2014ir}.
As in that work, we increase the Signal to Noise Ratio (SNR) by first averaging the reference spectrum by 16 channels. Then, adopting the baseline procedure of \citet{Leroy:2008jk}, we blanked all channels except for those within 400 \kms\ of either side of the emission peak. We then fit the unblanked channels with a first order baseline model and subtracted this model from all channels. Finally, we smoothed the spectrum to a resolution of 7.2\kms.

We show the single dish \HI\ spectrum in the top panel of \autoref{fig:spectrum} centered near the systematic velocity. This Figure also illustrates our fitting and measurement routines. We fit a single gaussian component both to guide the reader's eye, and to define the area over which to integrate the galaxy spectrum (gray line in \autoref{fig:spectrum}). Next we fit linear components to the sides of the emission (red lines) and take the width between these at 50\% of the maximum value as the linewidth (blue line). The halfway point of this line is the galaxy centeroid (green line). This procedure was originally adopted from \citet{Springob:2005db} and discussed in more detail in \citet{Pardy:2014ir}. \citet{Pardy:2014ir} also details our procedure for calculating measurement uncertainties. In brief, we use a conservative 10\% uncertainty for baseline and calibration results, to which we add an uncertainty based on the width and noise of the spectrum:
\begin{equation}
\epsilon = rms\sqrt{W_{50}\times\Delta V},
\end{equation}
where `$rms$' is the root mean squared noise of line-free channels, $W_{50}$ is the width at 50\% of the peak, and $\Delta V$ is the velocity resolution. 

We use the single dish \HI\ spectrum to measure the total \HI\ flux, and convert this to a mass using:
\begin{equation}
\frac{M_{HI}}{M_{\odot}} \approx 2.36\times10^5\left(\frac{D}{Mpc}\right)^2\int\left(\frac{S(\nu)}{Jy}\right)\left(\frac{d\nu}{km s^{-1}}\right).
\end{equation}
Where D is the distance and S is the flux density. In this work we take the previous optical line center of 6180 \kms\ \citep{Bergvall:2006ib} and derive a Hubble-flow distance of 88 Mpc using the cosmology from \citet{2011ApJS..192...18K}. This conversion from flux to mass assumes that the emission is optically thin, but does not require any knowledge of the spin temperature as is required by absorption-based masses. Using the derived mass and the known stellar mass we can then measure the fraction of neutral gas in the galaxy. 

To ensure that this result was not due to the presence of contamination within the GBT beam, we checked the NASA Extragalactic Database (NED) for known galaxies within 30 arcminutes of Haro 11. No companions were found.

\section{Results}
\label{sec:results}

    \begin{table}[htbp]
   \centering
   \caption{\HI\ and stellar properties for Haro 11.}
   \begin{threeparttable}
    \begin{tabular}{@{} llr @{}} 
      \multicolumn{2}{l}{Stellar Properties}\\
      \midrule
    	Property & Value & Source\\     
      \midrule
      V$_{opt}$  & 6180 \kms & \citep{Bergvall:2006ib} \\
      Distance & 88 Mpc & \citep{Bergvall:2006ib}\tnote{a}\\
      f$_{esc}$\tnote{b}      & 0.037 &  \citep{Ostlin:2009jp}\\
      EW\tnote{c} &  15.6 \AA & \citep{Ostlin:2009jp}\\
       Lum$_{\lya}$. & 8.4$\times$10$^{41}$erg s$^{-1}$ &  \citep{Ostlin:2009jp}\\
       SFR\tnote{d}         & 24 \msun yr$^{-1}$   &  \citep{Hayes:2007hk} \\
       M$_{\star}$ & 9.4$^{+12}_{-3.3}\times10^9$ \msun & \citep{Ostlin:2001ch} \\
       M$_{HII}$ & 10 \pom\ 1 $\times 10^8$ \msun & \citep{Bergvall:2002ea} \\
       L$_{B, \odot}$ & 1.8$\times$10$^{10}$L$_{\odot}$ & \citep{Bergvall:2002ea}\tnote{e}\\
      \midrule
     \multicolumn{2}{l}{\HI\ Properties (this work)} \\
     \midrule
     	Property &  \multicolumn{2}{l}{Value} \\     
       \midrule
    Flux &  \multicolumn{2}{l}{0.28\pom\ 0.04 Jy \kms} \\
      V$_{sys}$\tnote{f} & \multicolumn{2}{l}{6236 \pom\ 11 \kms} \\
      M$_{H I}$ & \multicolumn{2}{l}{5.1\pom0.7$\times 10^8$\msun} \\
      W$_{50}$ & \multicolumn{2}{l}{77\pom\ 21 \kms} \\
      W$_{20}$ & \multicolumn{2}{l}{125\pom\ 33 \kms} \\
      W$_{max}$\tnote{g}  & \multicolumn{2}{l}{163\pom\ 42 \kms} \\
      f$_{gas}$ & \multicolumn{2}{l}{ 0.05$^{+0.4}_{-0.4}$} \\
      \bottomrule
    \end{tabular}
        \begin{tablenotes}
        	   \item[a] Using H$_0$ = 70 \kms\ Mpc$^{-1}$.  \item[b] Escape fraction of \lya.
            \item[c] Equivalent width of \lya.  \item[d] \halpha\ derived star formation rate.
            \item[e] Using distance = 88 Mpc.  \item[f] Heliocentric radial velocity.
            \item[g] Full width of spectral line at zero-point crossing.
        \end{tablenotes}
     \end{threeparttable}
    \label{tab:params}
    \end{table}

We measured the total flux of Haro 11 as 0.28\pom 0.04 Jy \kms\ at a velocity center of 6237\pom11 \kms\ (using a heliocentric frame of rest). The un-smoothed spectrum has an rms noise of 3$\times$10$^{-3}$ Jy and a peak signal to noise ratio of 8. This line center corresponds to cosmological redshift distance of 89 \pom 2 Mpc. This central velocity is formally inconsistent both with the observed line center of 6179 \pom\ 16 \kms\ found by \citet{MacHattie:2014ipa} and with the \halpha\ line center of 6146 \kms\ from \citet{James:2013ia}. 

To test the validity of this detection, we split the dataset into the two separate polarizations and performed the analysis on each piece independently. We recover similar measurements from the two polarizations, all within our adopted uncertainties.

We compare the emission and absorption components in \autoref{fig:spectrum}. The absorption feature was unresolved in the observations of \citet{MacHattie:2014ipa}, and comes from a region of size $<$3.54 kpc near the center of Haro 11. 

\begin{figure*}[t] 
   \centering
   \includegraphics[width=7in]{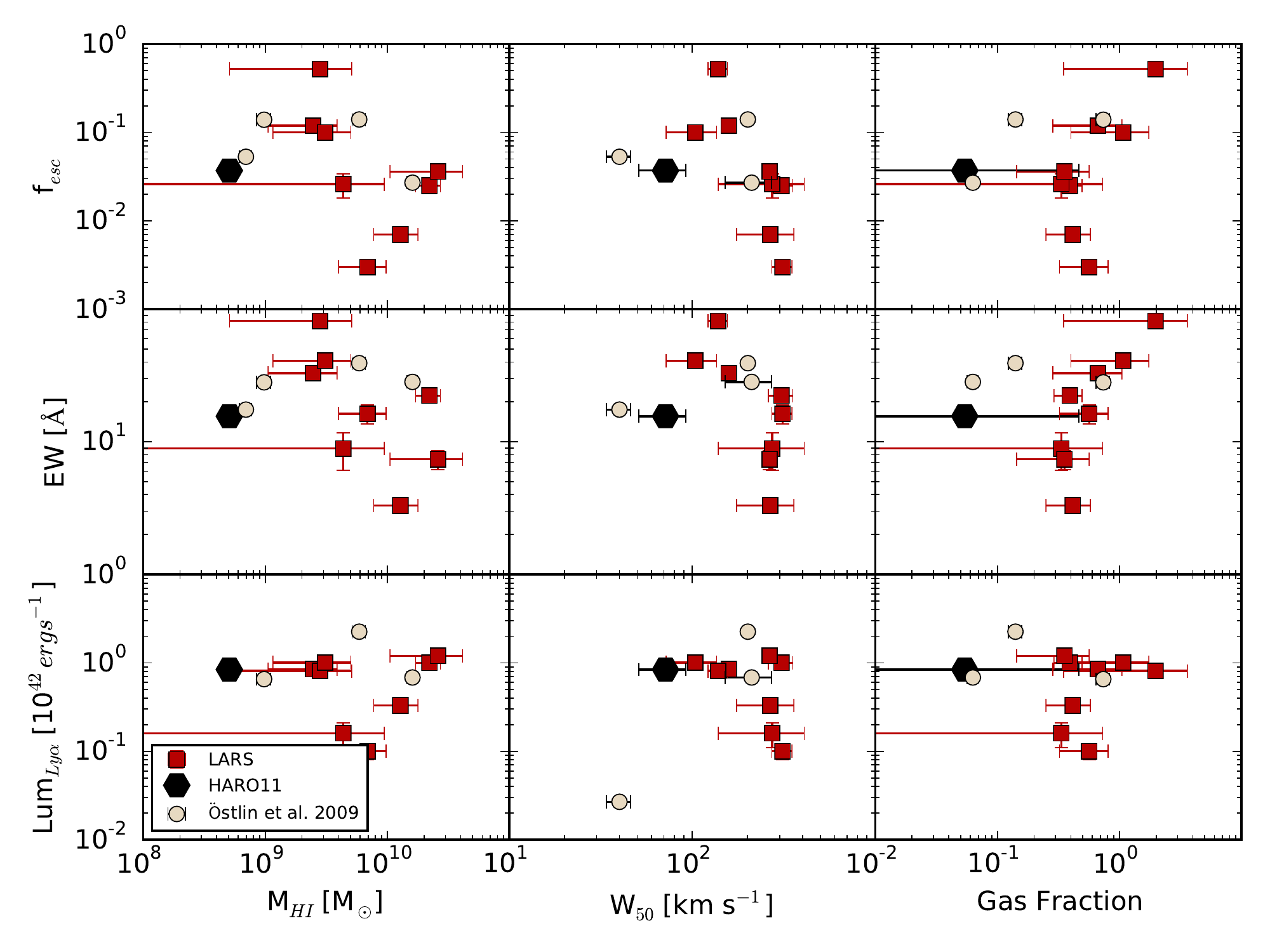} 
   \caption{\lya\ and \HI\ properties of Haro 11 compared with other nearby starburst galaxies. In all panels Haro 11 is shown as a black hexagon, the LARS galaxies with clean detections are red squares, and the local starburst galaxies (see \autoref{tab:lyaparams}) are gray circles. Only galaxies with a positive \lya\ luminosity are shown. Because of these quality and \lya\ luminosity cuts, we do not show LARS 4, 5, 6, 12, 13, and 14. Each column shows a different \HI\ property. \emph{Left column}: \HI\ mass. \emph{Middle column}: linewidth at 50\% of the line maximum. Note: ESO 338-04 is excluded from this panel because it has no published W$_{50}$ value. \emph{Right column}: gas fraction measured as M$_{HI}$/M$_{\star}$. Note: Tol65 is excluded from this panel because it has no published M$_{\star}$ value. The rows show different \lya\ properties. \emph{Top row}: the escape fraction of \lya. \emph{Middle row}: the equivalent width of \lya. \emph{Bottom row}: Luminosity of \lya}
   \label{fig:lyaprops}
\end{figure*}

Using the total flux and a distance of 88 Mpc, we found a total of 5.1 \pom 0.7 $\times$10$^8$ \msun\ of \HI\ gas in Haro 11, which makes it an order of magnitude deficient compared to LARS galaxies with similar \lya\ properties (see the left column of \autoref{fig:lyaprops}), but in line with other starburst galaxies from \citet{Ostlin:2009jp}. This \HI\ mass is less than half the upper limit provided by \citet{MacHattie:2014ipa} and is $\sim$5 times the order-of-magnitude limit provided by \citet{Bergvall:2006ib}. Due to the low \HI\ mass, Haro 11 has a gas fraction (M$_{HI}$/M$_{\star}$) of 0.05 \pom 0.4, the lowest of any of the local comparison galaxies for which \HI\ was detected (see right column of \autoref{fig:lyaprops} and \autoref{LYAProps} for details). 

We measure the velocity width at 50\% of the peak emission as 77 \pom 21 \kms\ and the velocity width at 20\% of the peak emission as 125 \pom\ 32 \kms. Haro 11 has smaller linewidth than any LARS galaxies, as seen in the middle column of \autoref{fig:lyaprops}. Our linewidth is narrower than the absorption derived \HI\ linewidth of 112 \pom 16 \kms\ \citep{MacHattie:2014ipa}. The total velocity extent is 163\pom 42 \kms, formally consistent with those of \citet{MacHattie:2014ipa} and \citet{Ostlin:1999cl} (192 and 190 \kms\ respectively), and smaller than the \Ha\ derived value of 283 \kms\ from \citet{James:2013ia}.

Although Haro 11 appears to have less \HI\ and a smaller linewidth when compared to galaxies in LARS with similar UV properties, none of these relationships, with the exception of the relationship between f$_{esc}$ and linewidth, are significant. We quantify this by measuring Spearman's $\rho$ coefficient \citep{Spearman:1904ic}, finding f$_{esc}$ and linewidth with a correlation coefficient of -0.76 (p-value of 0.005). We next measured the strength of the correlation before and after including Haro 11 and looked for instances where the two properties became less correlated after the inclusion of Haro 11. In all cases, except for the relationship between gas fraction and EW\footnote{Although still not significant after the inclusion. Before including Haro 11: $\rho$ = 0.40 (p-value = 0.22). After: $\rho$ = 0.45 (p-value = 0.15).}, the two variables became less correlated after including Haro 11. 

\section{Discussion}
\label{sec:discussion}

\subsection{\lya\ Properties}
\label{LYAProps}

We compare our observed \HI\ mass, linewidth, and gas fraction for Haro 11  to the other starbursts from \citet{Ostlin:2009jp} and with galaxies from LARS \citep{Hayes:2014jv}. Literature data for the \HI\ properties of the \citet{Ostlin:2009jp} sample was compiled from a variety of sources and is presented in \autoref{tab:lyaparams}. 
We show all results in \autoref{fig:lyaprops}. The rows show different \lya\ properties
\begin{itemize}
\item \textbf{f$_{esc}$.} The \lya\ escape fraction measured as the ratio of observed to intrinsic \lya\ luminosity \citep{Hayes:2005ew, 2013ApJ...765L..27H} which was found in \citet{Pardy:2014ir} to anti-correlate with total \HI\ mass ($\rho$ = -0.62). Shown in the top row of \autoref{fig:lyaprops}.
\item \textbf{EW.} The equivalent width of \lya\ which was also found in \citet{Pardy:2014ir} to anti-correlate with total \HI\ mass ($\rho$ = -0.57). Shown in the middle row of \autoref{fig:lyaprops}.
\item \textbf{Lum$_{\lya}$.} The Luminosity of \lya\ from \citet{Hayes:2007hk, Hayes:2014jv}. Shown in the bottom row of \autoref{fig:lyaprops}.
\end{itemize}

Haro 11 does not follow the correlations seen in \citet{Pardy:2014ir} for \HI\ mass with f$_{esc}$ (top left panel in \autoref{fig:lyaprops}) and with EW (middle left panel in \autoref{fig:lyaprops}), but one other starburst galaxy (Tol 65) is also nearly as offset as Haro 11 from the general trends. Haro 11 appears to also have a smaller linewidth (W$_{50}$) than LARS galaxies with similar f$_{esc}$ and EW values, but this difference appears less pronounced than the offset with respect to \HI\ mass. As discussed in \autoref{conclusions}, interactions are thought to facilitate the escape of \lya\ photons, and often produce asymmetric line profiles and increased luminosities compared with isolated systems \citep{Zaritsky:1997ev}. Given that most of the LARS galaxies also showed signs of interactions, this may partly explain why the correlations between UV properties and \HI\ linewidth were generally weaker than those with \HI\ mass, and why Haro 11 falls within the linewidth scatter even with a dearth of \HI.

In addition, the gas fraction seen in Haro 11 is nearly an order of magnitude smaller than any LARS galaxies as predicted by each of these three \lya\ properties. The galaxies in \citet{Ostlin:2009jp} are generally smaller, with lower \HI\ masses than the LARS sample (5.9$\times10^{9}$\msun\ for \citep{Ostlin:2009jp} and 1.6$\times10^{11}$\msun\ for LARS) . Because of this, these \lya\ emitters appear as extreme as Haro 11 in these relationships. In particular, two of the galaxies IRAS 08339+6517 and NGC 6090 have lower gas fractions than the LARS galaxies (although they have masses and line-widths consistent with the larger sample; see the right column of \autoref{fig:lyaprops}). Given the uncertain nature of the dependence on \HI\ properties for the propagation of \lya\ photons, a larger systematic study of \HI\ properties in \lya\ emitters is warranted. One such project is currently underway with the expanded Lyman Alpha Reference Sample (eLARS), which will increase the LARS sample by 28 galaxies (Melinder et al. in prep).

\subsection{Relation to other Galaxies}
\label{context}

\begin{figure}[h] 
   \centering
   \includegraphics[width=3.5in]{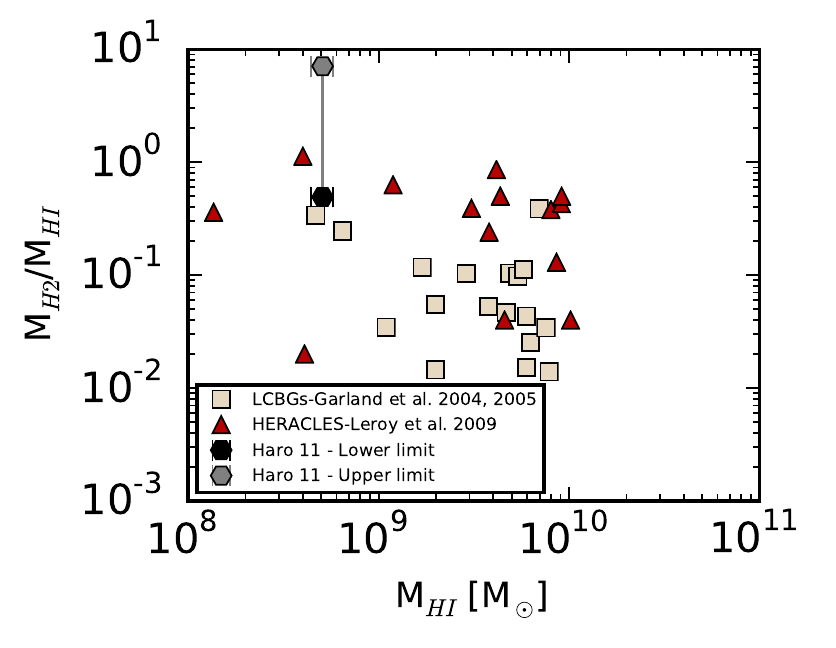} 
   \caption{Molecular gas fraction, M$_{H_2}$/M$_{HI}$, for nearby galaxies and Haro 11. Comparison data is from the HERACLES sample \citep{Leroy:2008jk} (red triangles)  and a sample of LCBGs from \citet{Garland:2004ec, Garland:2005bd} (gold squares). Haro 11 has an uncertain H$_2$ mass from \citet{Cormier:2014il}. According to the CO derived mass, using a galactic X$_{CO}$, Haro 11 has M$_{H_2}$ = 2.5$\times$10$^8$ \msun. The dust measurements, however, give an estimate of M$_{H_2}$ = 3.6$\times$10$^9$ \msun. We show both points as an upper and lower limit to the M$_{H_2}$/M$_{HI}$ ratio. At the lower limit, Haro 11 has the fifth highest molecular gas fraction in this sample. At the higher limit it has a factor of seven higher ratio than any HERACLES galaxies. }
   \label{fig:MH2MHI}
\end{figure}

\begin{figure*}[t] 
   \centering
   \includegraphics[width=7in]{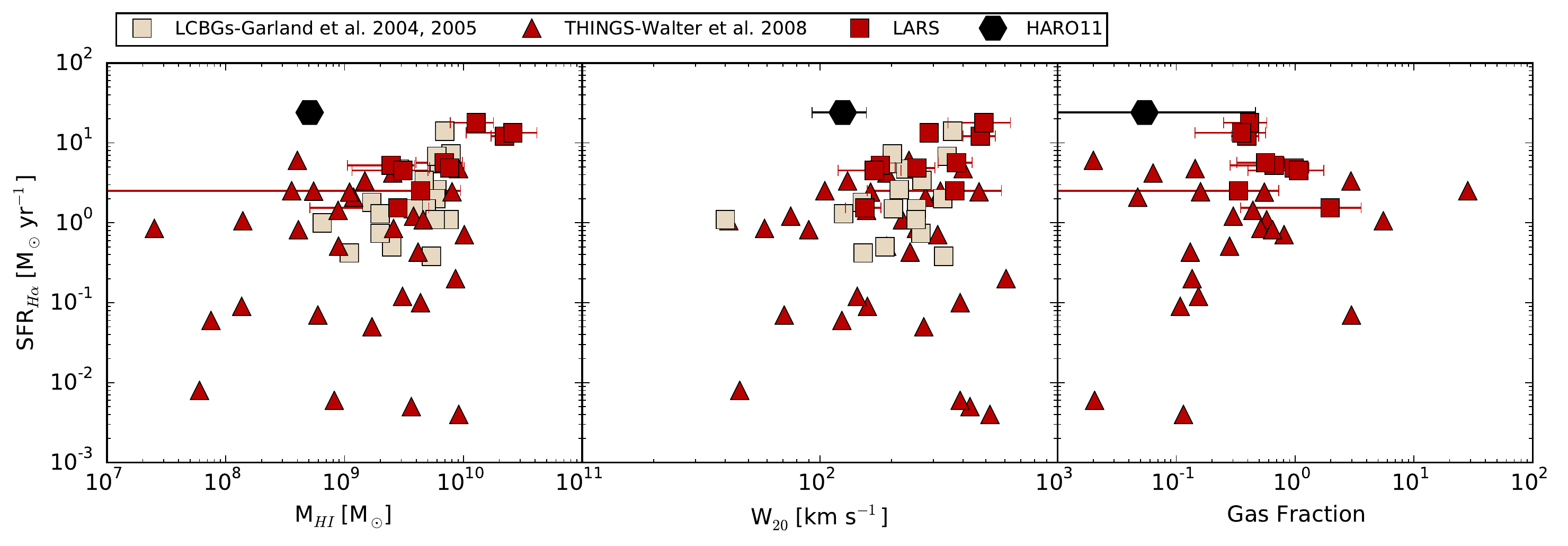} 
   \caption{SFR and \HI\ properties in Haro 11 and local galaxies. We compare with the LARS galaxies marked as clean detections (red squares), the THINGS galaxies \citep{Walter:2008bq} and other LCBGs from \citep{Garland:2004ec, Garland:2005bd} (red triangles and gold squares, respectively). Because of the quality cuts, we do not show LARS 5, 6, 12, 13, and 14. \emph{Left:} The \HI\ mass shows that Haro 11 has an elevated SFR compared to other galaxies. \emph{Middle:} The velocity width at 20\% of the peak (W$_{20}$) of Haro 11 is less than would be expected from other galaxies. \emph{Right:} The gas fraction (measured as M$_{HI}$/M$_{\star}$) matches expectations of high star forming galaxies from these two samples.}
   \label{fig:SFR}
\end{figure*}

Haro 11 has a stellar mass about one third that of the MW \citep{Ostlin:2001ch, Licquia:2015eb}, but a total mass near the upper limits of the LMC \citep{Ostlin:2015tc, Penarrubia:2015vk} and an \HI\ mass that is reminiscent of dwarfs. Because of this range of properties, we compare Haro 11 to a wide range of galaxy samples including regular star forming galaxies, dwarf galaxies, and galaxies selected for their similar blue colors or low metallicities as Haro 11. 

\citet{Cormier:2014il} measured the H$_2$ mass of Haro 11 using both CO and dust tracers. Haro 11 appears to have a small CO flux, and using Galaxy scaled X$_{CO}$ conversions gives M$_{H_2}$ = 2.5$\times$10$^8$ \msun, whereas dust measurements give an estimate of M$_{H_2}$ = 3.6$\times$10$^9$ \msun. Using these dust masses, a metallicity scaled X$_{co}$ provides a higher mass of M$_{H_2}$ = 2.5$\times$10$^9$ \msun. \citet{Cormier:2014il} finds similar discrepancies between the CO and dust-derived H$_2$ masses for other galaxies in the Herschel Dwarf Galaxy Survey. All of the galaxies in their sample have faint CO lines and X$_{CO}$ factors much larger than the Galactic value and therefore have uncertain H$_2$ masses. We can use this H$_2$ mass to derive an \HI\ to H$_2$ mass ratio. Because there is such a large discrepancy between different methods, we choose to use the full range of H$_2$ masses reported by \citet{Cormier:2014il} in the analysis. This gives a range of M$_{H_2}$/M$_{HI}$ = 0.43 - 6.2 depending on the mass of H$_2$ used. 

If Haro 11 is indeed the product of a recent merger \citep{Ostlin:2015tc}, then comparisons with non-interacting galaxies may be biased. If we take these results at face value, however, we find that at the lower limit, Haro 11 has the fifth highest molecular gas fraction in this sample. At the higher limit it has a factor of seven higher ratio than any galaxies from the Heterodyne Receiver Array CO Line Extragalactic Survey (HERACLES; \citealt{Leroy:2009di}). See \autoref{fig:MH2MHI}. This large M$_{H_2}$/M$_{HI}$  ratio is perhaps because infrared luminous galaxies are often seen with low atomic gas fractions \citep{Mirabel:1989ds}. 
 
In \autoref{fig:SFR} we compare the star formation and \HI\ properties of Haro 11 to Luminous Compact Blue Galaxies (LCBGs) from \citet{Garland:2004ec, Garland:2005bd}, to The \HI\ in Nearby Galaxies Survey (THINGS) galaxies \citep{Walter:2008bq}, and to LARS galaxies. The star formation rate for Haro 11 is measured from \Ha\ emission from \citet{Hayes:2007hk}. Haro 11 appears to be missing gas, both in terms of its low \HI\ mass compared to other galaxies with similar SFR (left panel in \autoref{fig:SFR}) and in terms of its narrow linewidth (middle panel). The gas fraction, however, paints a different picture. Haro 11 matches a trend of lower gas fractions toward higher star formation rates found in both the THINGS galaxies and the LBCGs (right panel of \autoref{fig:SFR}). Taken together, these point to Haro 11 not as missing atomic gas, but as a standard low-mass galaxy that is undergoing an unusually efficient star formation episode. With an instantaneous \Ha\ SFR of 24 \msun yr$^{-1}$ \citep{Hayes:2007hk}, and the range of gas masses provided above (including a correction for primordial He), the gas will be depleted in 0.04-0.2 Gyrs. 

\begin{figure}[h] 
   \centering
   \includegraphics[width=3.5in]{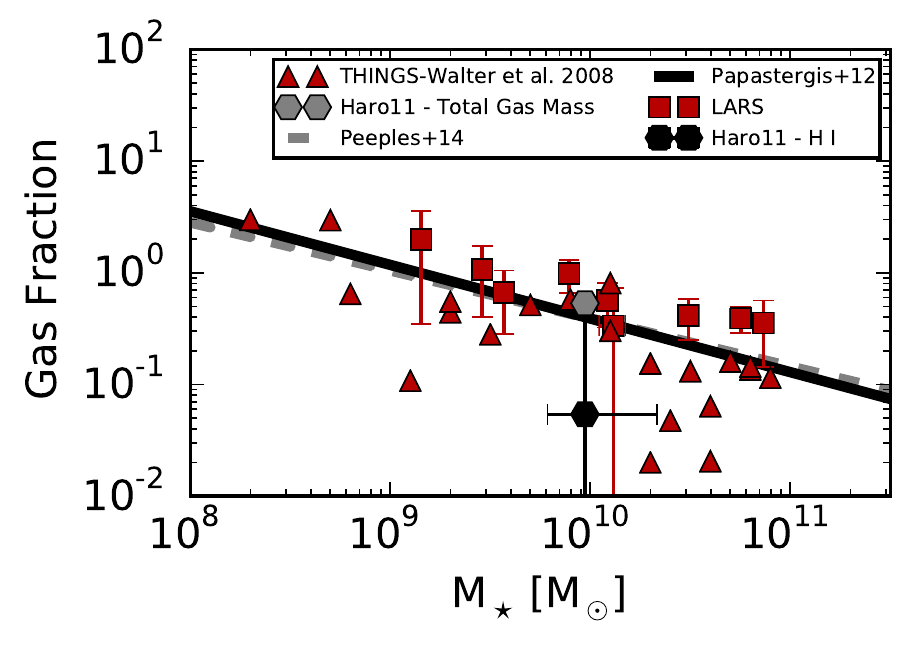} 
   \caption{The relation between stellar mass and the gas mass fraction. LARS galaxies \citep{Pardy:2014ir} are converted from Salpeter to Chabrier IMF and are shown as red squares. THINGS galaxies \citep{Walter:2008bq} are shown as red triangles. We compare these galaxies and Haro 11, shown as a black hexagon and using a gas fraction derived only from the \HI, to the relations found by \citet{Peeples:2014fe} and \citet{Papastergis:2012cb}. The detections in the LARS sample, and many galaxies from THINGS, follow these predictions, but Haro 11 and four of the THINGS galaxies (IC 2574, NGC 3621, NGC 4449, and NGC 5055) lie an order of magnitude below the predicted gas fraction. The gray point shows the total gas fraction of Haro 11 (derived using the combined \HI, H$_2$, and \halpha\ masses) compared with its stellar mass. This follows the predictions found in \citet{Peeples:2014fe}.}
   \label{fig:fgas_mstar}
\end{figure}

Data from ALFALFA \citep{Giovanelli:2005jt}, and HERACLES \citep{Leroy:2008jk}, among others, have been used to demonstrate relationships between \HI\ gas fraction and stellar mass (\citealt{Papastergis:2012cb}; log $F_g = -0.48\, \textrm{log }M_{\star}+4.39$) and between \HI\ + H$_2$ gas fraction and stellar mass (\citealt{Peeples:2014fe}; log $F_g = -0.48\, \textrm{log }M_{\star}+4.39$) for normal star-forming disk galaxies. The general trend is that galaxies with smaller total stellar mass have a higher gas fraction. The \HI\ mass in Haro 11 does not follow this prediction and, as seen in \autoref{fig:fgas_mstar}, lies more than an order of magnitude below the trend. Haro 11 is interesting in this respect for two reasons. First, the \HII\ mass is roughly twice the \HI\ mass.  Second, Haro 11 has a large H$_2$/\HI\ fraction. After assuming an upper limit of H$_2$ mass of 3.6$\times$10$^9$\msun, and 10$\times$10$^8$\msun\ of ionized hydrogen, we can place Haro 11 on the \HI\ + H$_2$ - stellar mass relationship found in \citet{Peeples:2014fe}.

\begin{figure*}[t] 
   \centering
   \includegraphics[width=7in]{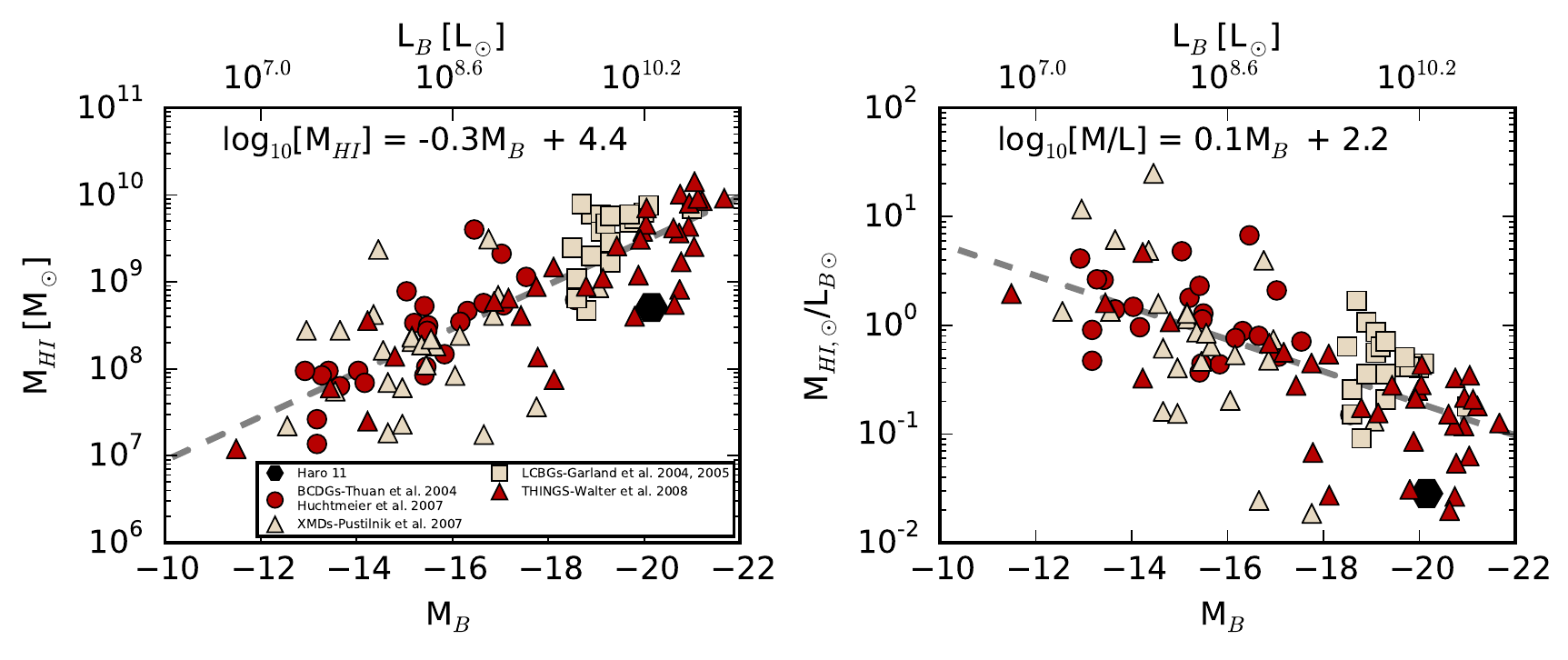} 
   \caption{Comparison of \HI\ mass with B band luminosity and magnitude. Comparison data in both panels is from BCDGs \citep{Thuan:2004hq, Huchtmeier:2007gk} (red circles), XMDs \citep{Pustilnik:2007el} (gold triangles), LCBGs \citep{Garland:2004ec, Garland:2005bd} (gold squares) and THINGS \citep{Walter:2008bq} (red triangles). \emph{Left:} The M$_{HI}$ with respect to M$_B$. We also include the transformed L$_{B}$ in solar luminosities on the upper axis, assuming a B band magnitude of the sun of 5.48. Haro 11 is below the main trend, but sits with a few other THINGS galaxies. \emph{Right:} The M$_{HI}$/L$_B$ with respect to M$_B$. Although Haro 11 has a very low gas fraction compared to other BCDGs, LCBGs, and XMDs, it follows the general trend of decreasing mass-to-light ratio with increasing luminosity. }
   \label{fig:MHILB}
\end{figure*}

\citet{Smoker:2000ug} finds a trend in M$_{HI}$/L$_B$, with higher luminosity galaxies having lower gas ratios. We observe this trend in samples of BCDGs from \citet{Thuan:2004hq} and \citet{Huchtmeier:2007gk}, a sample of LCBGs from \citet{Garland:2004ec, Garland:2005bd}, a sample of extreme metal deficient dwarfs \citep[XMDs;][]{Pustilnik:2007el}, and with galaxies from THINGS \citep{Walter:2008bq}. All galaxies from these papers have been corrected for our assumed cosmology wherever possible.  \autoref{fig:MHILB} shows how these galaxies match with the observed luminosity and \HI\ mass of Haro 11. The right panel of \autoref{fig:MHILB} shows the classic relationship of M$_{HI}$/L$_B$ with M$_B$ found in other papers, while the left panel shows a similar correlation of M$_{HI}$ with respect to M$_{B}$. In both panels we include the transformation to luminosity in the top axis, assuming a solar B band magnitude of 5.48. The two relationships are highly correlated. The left and right panels have a Spearman's correlation coefficient of -0.79 and 0.7 respectively and can be well described by simple power law fits (log$_{10}$M$_{HI}$=-0.3M$_B$ + 4.4 and log$_{10}$M/L=-0.1M$_B$ + 2.2). Haro 11 is a $\sim$2 sigma outlier from the general trends (each relationship shows scatter of 0.46 dex), but is not the most extreme outlier.  

\subsection{Kinematics}
\label{kinematics}

The ISM kinematics of Haro 11 are highly irregular. The SiII line, seen in absorption by \citet{1998astro.ph..9096K}, is blue shifted by 58 \kms\ from the optical line center. \citet{Sandberg:2013hw} sees large velocity offsets between the ionized (\halpha) and neutral gas (using a weak Na doublet line) that vary across the star forming knots. In Knot B, the neutral gas is blueshifted by 44 \kms, while in Knot C the neutral gas is redshifted by 32 \kms. \citet{Heckman:2015dq} finds an outflow of 160 \kms\ using interstellar absorption lines in the warm ionized gas. Adding to this confusion, our \HI\ emission line is redshifted nearly 60 \kms\ compared to the \HI\ absorption \citep{MacHattie:2014ipa} or \halpha\ emission \citep{James:2013ia} emission. 

It is possible that we are seeing evidence of outflows carving cavities for the \lya\ propagation. Such a phenomenon is seen in other irregular galaxies such as NGC 2366, which features two prominent outflows \citep{vanEymeren:2009dh}. Although \halpha\ and \HI\ kinematics are aligned across much of NGC 2366, the red-shifted \halpha\ outflow is in a hole in the neutral gas and has no \HI\ counterpart. 

Alternatively, the \HI\ line in Haro 11 might be shifted due to large scale effects such as tidal tails. Such a tidal arm was proposed by \citet{Ostlin:2001ch, Ostlin:2015tc} to explain a redshifted \HII\ component. Although the measured \HI\ velocity width is $\sim$ 2.5 times narrower than the FWHM of the ionized gas, the ionized gas is multicomponent \citep{Ostlin:2001ch}. In particular \citet{Ostlin:2015tc} found a redshifted component with the same width ($\sigma$=30-25 km/s) as for the ionized hydrogen. They proposed that this component is a tidal arm. The \HI\ that we see in emission could possibly be associated with this component, in which case it would not be associated with the 
main star formation site in the center (which is associated with the HI gas seen in 
absorption since the velocity matches and the width is consistent). 

The total \HI\  velocity extent is 160 \kms, which suggests another possibility. The outflowing gas we see might be right on the edge of the \HI\ distribution and the bulk of this \HI\ material might be simply behind the \HII\ regions. 

\section{Conclusions}
\label{conclusions}

Haro 11 is a local starburst galaxy undergoing a major merger of two dwarf galaxies \citep{Ostlin:1999cl, Ostlin:2015tc}. It has long been known to be a local \lya-emitter, but until now has only had upper limits placed on its cold gas content. Our main contribution is the first robust detection of \HI\ gas in emission from this blue compact galaxy consistent with previous upper limits \citep{Bergvall:2000tg, MacHattie:2014ipa}. We find a dearth of \HI\ gas and a linewidth at 50\% of maximum that is smaller relative to other local starburst and LARS galaxies \citep{Ostlin:2014vk, Hayes:2014jv, Pardy:2014ir}.

Given the small \HI\ mass, Haro 11 has an elevated M$_{H2}$/M$_{HI}$ ratio and a very low gas fraction compared to most local galaxies. Much of the hydrogen that remains has been heated during the merger process. \citet{Bergvall:2002ea} derived a mass of 10 \pom\ 1 $\times$10$^8$\msun\ for the ionized hydrogen, twice as much as in neutral hydrogen.  
Haro 11 is undergoing an unusually efficient star formation episode, and matches a trend of lower gas fractions toward higher star formation rates and shares \HI\ properties with galaxies of similar B-band magnitude.

Haro 11 is a known merging galaxy, which could explain the presence of tidal arms and complicated ISM kinematics. Two other merging local \lya-emitters ESO\,338$-$IG\,004 (Tol\,1924$-$416) and IRAS 08339+6517 were studied by \citet{2004ApJ...608..768C} who found extended neutral gas connecting the main galaxy and companions. The processes that govern \lya\ transport are complicated, but most often thought to involve the bulk motion of \HI\ gas \citep{Hayes:2015wy}. Mapping the \HI\ gas in emission across Haro 11 will be the only way to test these hypotheses, but this task remains a challenge for the current generation of radio telescopes. As we move to the high resolution \HI\ era with the Square Kilometer Array (SKA), Haro 11 will be an excellent nearby target to probe the complicated coupling to cold neutral gas and \lya\ propagation. Probing the three distinct star formation knots will reveal processes at work in enabling \lya\ escape.

\acknowledgements

This research made use of Astropy\footnote{http://www.astropy.org/}, a community-developed core Python
package for Astronomy (Astropy Collaboration, 2013). This research has made
use of the NASA/IPAC Extragalactic Database (NED), which is operated
by the Jet Propulsion Laboratory, California Institute of Technology,
under contract with the National Aeronautics and Space Administration,
and NASA's Astrophysics Data System. SP thanks Jeremy MacHattie for providing the data for his figure 2 and for the discussion about his reduction methods. M.H. and G.\"O. acknowledge the support of the Swedish Research Council (Vetenskapsr{\aa}det) and the Swedish National Space Board (SNSB). M.H. is a Fellow of the Knut and Alice Wallenberg Foundation. 

\bibliography{Haro11}

\begin{thebibliography}{58}
\expandafter\ifx\csname natexlab\endcsname\relax\def\natexlab#1{#1}\fi

\bibitem[{Adamo {et~al.}(2010)Adamo, {\"O}stlin, Zackrisson, Hayes, Cumming, \&
  Micheva}]{Adamo:2010jv}
Adamo, A., {\"O}stlin, G., Zackrisson, E., et~al. 2010, MNRAS, 407, 870

\bibitem[{Atek {et~al.}(2008)Atek, Kunth, Hayes, {\"O}stlin, \&
  Mas-Hesse}]{Atek:2008hn}
Atek, H., Kunth, D., Hayes, M., et~al. 2008, A{\&}A, 488, 491

\bibitem[{Atek {et~al.}(2009)Atek, Kunth, Schaerer, Hayes, Deharveng,
  {\"O}stlin, \& Mas-Hesse}]{Atek:2009hra}
Atek, H., Kunth, D., Schaerer, D., et~al. 2009, A{\&}A, 506, L1

\bibitem[{Bekki(2008)}]{Bekki:2008jt}
Bekki, K. 2008, MNRAS, 388, L10

\bibitem[{Bergvall {et~al.}(2000)Bergvall, Masegosa, {\"O}stlin, \&
  Cernicharo}]{Bergvall:2000tg}
Bergvall, N., Masegosa, J., {\"O}stlin, G., et~al. 2000, A{\&}A, 359, 41

\bibitem[{Bergvall \& {\"O}stlin(2002)}]{Bergvall:2002ea}
Bergvall, N. and {\"O}stlin, G. 2002, A{\&}A, 390, 891

\bibitem[{Bergvall {et~al.}(2006)Bergvall, Zackrisson, Andersson, Arnberg,
  Masegosa, \& {\"O}stlin}]{Bergvall:2006ib}
Bergvall, N., Zackrisson, E., Andersson, B.~G., et~al. 2006, A{\&}A, 448, 513

\bibitem[{Cannon {et~al.}(2004)Cannon, Skillman, Kunth, Leitherer, Mas-Hesse,
  {\"O}stlin, \& Petrosian}]{2004ApJ...608..768C}
Cannon, J.~M., Skillman, E.~D., Kunth, D., et~al. 2004, ApJ, 608, 768

\bibitem[{Cormier {et~al.}(2014)Cormier, Madden, Lebouteiller, Hony, Aalto,
  Costagliola, Hughes, R{\'e}my-Ruyer, Abel, Bayet, Bigiel, Cannon, Cumming,
  Galametz, Galliano, Viti, \& Wu}]{Cormier:2014il}
Cormier, D., Madden, S.~C., Lebouteiller, V., et~al. 2014, A{\&}A, 564, 121

\bibitem[{Garland {et~al.}(2004)Garland, Pisano, Williams, Guzman, \&
  Castander}]{Garland:2004ec}
Garland, C.~A., Pisano, D.~J., Williams, J.~P., et~al. 2004, ApJ, 615, 689

\bibitem[{Garland {et~al.}(2005)Garland, Williams, Pisano, Guzman, Castander,
  \& Brinkmann}]{Garland:2005bd}
Garland, C.~A., Williams, J.~P., Pisano, D.~J., et~al. 2005, ApJ, 624, 714

\bibitem[{Giavalisco {et~al.}(1996)Giavalisco, Koratkar, \&
  Calzetti}]{1996ApJ...466..831G}
Giavalisco, M., Koratkar, A., and Calzetti, D. 1996, Astrophysical Journal
  v.466, 466, 831

\bibitem[{Giovanelli {et~al.}(2005)Giovanelli, Haynes, Kent, Perillat,
  Saintonge, Brosch, Catinella, Hoffman, Stierwalt, Spekkens, Lerner, Masters,
  Momjian, Rosenberg, Springob, Boselli, Charmandaris, Darling, Davies, Lambas,
  Gavazzi, Giovanardi, Hardy, Hunt, Iovino, Karachentsev, Karachentseva,
  Koopmann, Marinoni, Minchin, MULLER, Putman, Pantoja, Salzer, Scodeggio,
  Skillman, Solanes, Valotto, van Driel, \& van Zee}]{Giovanelli:2005jt}
Giovanelli, R., Haynes, M.~P., Kent, B.~R., et~al. 2005, AJ, 130, 2598

\bibitem[{Guaita {et~al.}(2015)Guaita, Melinder, Hayes, {\"O}stlin, Gonzalez,
  Micheva, Adamo, Mas-Hesse, Sandberg, Ot{\'\i}-Floranes, Schaerer, Verhamme,
  Freeland, Orlitov{\'a}, Laursen, Cannon, Duval, Rivera-Thorsen, Herenz,
  Kunth, Atek, Puschnig, Gruyters, \& Pardy}]{Guaita:2015kr}
Guaita, L., Melinder, J., Hayes, M., et~al. 2015, A{\&}A, 576, A51

\bibitem[{Hayes(2015)}]{Hayes:2015wy}
Hayes, M. 2015, Publ. Astron. Soc. Aust, 32, e027

\bibitem[{Hayes {et~al.}(2007)Hayes, {\"O}stlin, Atek, Kunth, Mas-Hesse,
  Leitherer, Jimenez-Bailon, \& Adamo}]{Hayes:2007hk}
Hayes, M., {\"O}stlin, G., Atek, H., et~al. 2007, MNRAS, 382, 1465

\bibitem[{Hayes {et~al.}(2014)Hayes, {\"O}stlin, Duval, Sandberg, Guaita,
  Melinder, Adamo, Schaerer, Verhamme, Orlitov{\'a}, Mas-Hesse, Cannon, Atek,
  Kunth, Laursen, Ot{\'\i}-Floranes, Pardy, Rivera-Thorsen, \&
  Herenz}]{Hayes:2014jv}
Hayes, M., {\"O}stlin, G., Duval, F., et~al. 2014, ApJ, 782, 6

\bibitem[{Hayes {et~al.}(2005)Hayes, {\"O}stlin, Mas-Hesse, Kunth, Leitherer,
  \& Petrosian}]{Hayes:2005ew}
Hayes, M., {\"O}stlin, G., Mas-Hesse, J.~M., et~al. 2005, A{\&}A, 438, 71

\bibitem[{Hayes {et~al.}(2013)Hayes, {\"O}stlin, Schaerer, Verhamme, Mas-Hesse,
  Adamo, Atek, Cannon, Duval, Guaita, Herenz, Kunth, Laursen, Melinder,
  Orlitov{\'a}, Ot{\'\i}-Floranes, \& Sandberg}]{2013ApJ...765L..27H}
Hayes, M., {\"O}stlin, G., Schaerer, D., et~al. 2013, The Astrophysical Journal
  Letters, 765, L27

\bibitem[{Heckman {et~al.}(2015)Heckman, Alexandroff, Borthakur, Overzier, \&
  Leitherer}]{Heckman:2015dq}
Heckman, T.~M., Alexandroff, R.~M., Borthakur, S., et~al. 2015, ApJ, 809, 147

\bibitem[{Howell {et~al.}(2010)Howell, Armus, Mazzarella, Evans, Surace,
  Sanders, Petric, Appleton, Bothun, Bridge, Chan, Charmandaris, Frayer, Haan,
  Inami, Kim, Lord, Madore, Melbourne, Schulz, U, Vavilkin, Veilleux, \&
  Xu}]{Howell:2010ib}
Howell, J.~H., Armus, L., Mazzarella, J.~M., et~al. 2010, ApJ, 715, 572

\bibitem[{Huchtmeier {et~al.}(2007)Huchtmeier, Petrosian, {Gopal-Krishna}, \&
  Kunth}]{Huchtmeier:2007gk}
Huchtmeier, W.~K., Petrosian, A., {Gopal-Krishna}, et~al. 2007, A{\&}A, 462,
  919

\bibitem[{James {et~al.}(2013)James, Tsamis, Walsh, Barlow, \&
  Westmoquette}]{James:2013ia}
James, B.~L., Tsamis, Y.~G., Walsh, J.~R., et~al. 2013, MNRAS, 430, 2097

\bibitem[{Komatsu {et~al.}(2011)Komatsu, Smith, Dunkley, Bennett, Gold,
  Hinshaw, Jarosik, Larson, Nolta, Page, Spergel, Halpern, Hill, Kogut, Limon,
  Meyer, Odegard, Tucker, Weiland, Wollack, \& Wright}]{2011ApJS..192...18K}
Komatsu, E., Smith, K.~M., Dunkley, J., et~al. 2011, The Astrophysical Journal
  Supplement, 192, 18

\bibitem[{Kunth {et~al.}(2003)Kunth, Leitherer, Mas-Hesse, {\"O}stlin, \&
  Petrosian}]{2003ApJ...597..263K}
Kunth, D., Leitherer, C., Mas-Hesse, J.~M., et~al. 2003, ApJ, 597, 263

\bibitem[{Kunth {et~al.}(1998{\natexlab{a}})Kunth, Mas-Hesse, Terlevich,
  Terlevich, Lequeux, \& Fall}]{1998A&A...334...11K}
Kunth, D., Mas-Hesse, J.~M., Terlevich, E., et~al. 1998{\natexlab{a}}, A{\&}A,
  334, 11

\bibitem[{Kunth {et~al.}(1998{\natexlab{b}})Kunth, Terlevich, Terlevich, \&
  Tenorio-Tagle}]{1998astro.ph..9096K}
Kunth, D., Terlevich, E., Terlevich, R., et~al. 1998{\natexlab{b}}, arXiv,
  9809096

\bibitem[{Leitet {et~al.}(2013)Leitet, Bergvall, Hayes, Linn{\'e}, \&
  Zackrisson}]{Leitet:2013cw}
Leitet, E., Bergvall, N., Hayes, M., et~al. 2013, A{\&}A, 553, 106

\bibitem[{Leroy {et~al.}(2009)Leroy, Walter, Bigiel, Usero, Weiss, Brinks,
  de~Blok, Kennicutt, Schuster, Kramer, Wiesemeyer, \& Roussel}]{Leroy:2009di}
Leroy, A.~K., Walter, F., Bigiel, F., et~al. 2009, AJ, 137, 4670

\bibitem[{Leroy {et~al.}(2008)Leroy, Walter, Brinks, Bigiel, de~Blok, Madore,
  \& Thornley}]{Leroy:2008jk}
Leroy, A.~K., Walter, F., Brinks, E., et~al. 2008, AJ, 136, 2782

\bibitem[{Licquia \& Newman(2015)}]{Licquia:2015eb}
Licquia, T.~C. and Newman, J.~A. 2015, ApJ, 806, 96

\bibitem[{MacHattie {et~al.}(2014)MacHattie, Irwin, Madden, Cormier, \&
  R{\'e}my-Ruyer}]{MacHattie:2014ipa}
MacHattie, J.~A., Irwin, J.~A., Madden, S.~C., et~al. 2014, MNRAS, 438, L66

\bibitem[{Martin {et~al.}(1991)Martin, Bottinelli, Gouguenheim, \&
  Dennefeld}]{1991A&A...245..393M}
Martin, J.~M., Bottinelli, L., Gouguenheim, L., et~al. 1991, A{\&}A, 245, 393

\bibitem[{Mirabel \& Sanders(1989)}]{Mirabel:1989ds}
Mirabel, I.~F. and Sanders, D.~B. 1989, ApJ, 340, L53

\bibitem[{{\"O}stlin {et~al.}(2001){\"O}stlin, Amram, Bergvall, Masegosa,
  Boulesteix, \& Marquez}]{Ostlin:2001ch}
{\"O}stlin, G., Amram, P., Bergvall, N., et~al. 2001, A{\&}A, 374, 800

\bibitem[{{\"O}stlin {et~al.}(1999){\"O}stlin, Amram, Masegosa, Bergvall, \&
  Boulesteix}]{Ostlin:1999cl}
{\"O}stlin, G., Amram, P., Masegosa, J., et~al. 1999, Astronomy and
  Astrophysics Supplement Series, 137, 419

\bibitem[{{\"O}stlin {et~al.}(2014){\"O}stlin, Hayes, Duval, Sandberg,
  Rivera-Thorsen, Marquart, Orlitov{\'a}, Adamo, Melinder, Guaita, Atek,
  Cannon, Gruyters, Herenz, Kunth, Laursen, Mas-Hesse, Micheva,
  Ot{\'\i}-Floranes, Pardy, Roth, Schaerer, \& Verhamme}]{Ostlin:2014vk}
{\"O}stlin, G., Hayes, M., Duval, F., et~al. 2014, ApJ, 797, 11

\bibitem[{{\"O}stlin {et~al.}(2009){\"O}stlin, Hayes, Kunth, Mas-Hesse,
  Leitherer, Petrosian, \& Atek}]{Ostlin:2009jp}
{\"O}stlin, G., Hayes, M., Kunth, D., et~al. 2009, AJ, 138, 923

\bibitem[{{\"O}stlin {et~al.}(2015){\"O}stlin, Marquart, Cumming, Fathi,
  Bergvall, Adamo, Amram, \& Hayes}]{Ostlin:2015tc}
{\"O}stlin, G., Marquart, T., Cumming, R.~J., et~al. 2015, A{\&}A, 583, A55

\bibitem[{Papastergis {et~al.}(2012)Papastergis, Cattaneo, Huang, Giovanelli,
  \& Haynes}]{Papastergis:2012cb}
Papastergis, E., Cattaneo, A., Huang, S., et~al. 2012, ApJ, 759, 138

\bibitem[{Pardy {et~al.}(2014)Pardy, Cannon, {\"O}stlin, Hayes, Rivera-Thorsen,
  Sandberg, Adamo, Freeland, Herenz, Guaita, Kunth, Laursen, Mas-Hesse,
  Melinder, Orlitov{\'a}, Ot{\'\i}-Floranes, Puschnig, Schaerer, \&
  Verhamme}]{Pardy:2014ir}
Pardy, S.~A., Cannon, J.~M., {\"O}stlin, G., et~al. 2014, ApJ, 794, 101

\bibitem[{Peeples {et~al.}(2014)Peeples, Werk, Tumlinson, Oppenheimer,
  Prochaska, Katz, \& Weinberg}]{Peeples:2014fe}
Peeples, M.~S., Werk, J.~K., Tumlinson, J., et~al. 2014, ApJ, 786, 54

\bibitem[{Pe{\~n}arrubia {et~al.}(2015)Pe{\~n}arrubia, G{\'o}mez, Besla, Erkal,
  \& Ma}]{Penarrubia:2015vk}
Pe{\~n}arrubia, J., G{\'o}mez, F.~A., Besla, G., et~al. 2015

\bibitem[{Pustilnik \& Martin(2007)}]{Pustilnik:2007el}
Pustilnik, S.~A. and Martin, J.~M. 2007, A{\&}A, 464, 859

\bibitem[{Rivera-Thorsen {et~al.}(2015)Rivera-Thorsen, Hayes, {\"O}stlin,
  Duval, Orlitov{\'a}, Verhamme, Mas-Hesse, Schaerer, Cannon,
  Ot{\'\i}-Floranes, Sandberg, Guaita, Adamo, Atek, Herenz, Kunth, Laursen, \&
  Melinder}]{RiveraThorsen:2015ct}
Rivera-Thorsen, T.~E., Hayes, M., {\"O}stlin, G., et~al. 2015, ApJ, 805, 14

\bibitem[{Sandberg {et~al.}(2013)Sandberg, {\"O}stlin, Hayes, Fathi, Schaerer,
  Mas-Hesse, \& Rivera-Thorsen}]{Sandberg:2013hw}
Sandberg, A., {\"O}stlin, G., Hayes, M., et~al. 2013, A{\&}A, 552, A95

\bibitem[{Smoker {et~al.}(2000)Smoker, Davies, Axon, \& Hummel}]{Smoker:2000ug}
Smoker, J.~V., Davies, R.~D., Axon, D.~J., et~al. 2000, A{\&}A, 361, 19

\bibitem[{Spearman(1904)}]{Spearman:1904ic}
Spearman, C. 1904, The American Journal of Psychology, 15, 72

\bibitem[{Springob {et~al.}(2005)Springob, Haynes, Giovanelli, \&
  Kent}]{Springob:2005db}
Springob, C.~M., Haynes, M.~P., Giovanelli, R., et~al. 2005, ApJS, 160, 149

\bibitem[{Tenorio-Tagle {et~al.}(1999)Tenorio-Tagle, Silich, Kunth, Terlevich,
  \& Terlevich}]{TenorioTagle:1999bx}
Tenorio-Tagle, G., Silich, S.~A., Kunth, D., et~al. 1999, MNRAS, 309, 332

\bibitem[{Thuan {et~al.}(2004)Thuan, Hibbard, \& L{\'e}vrier}]{Thuan:2004hq}
Thuan, T.~X., Hibbard, J.~E., and L{\'e}vrier, F. 2004, AJ, 128, 617

\bibitem[{Vader {et~al.}(1993)Vader, Frogel, Terndrup, \&
  Heisler}]{Vader:1993fk}
Vader, J.~P., Frogel, J.~A., Terndrup, D.~M., et~al. 1993, AJ, 106, 1743

\bibitem[{van Driel {et~al.}(2001)van Driel, Gao, \&
  Monnier-Ragaigne}]{vanDriel:2001gx}
van Driel, W., Gao, Y., and Monnier-Ragaigne, D. 2001, A{\&}A, 368, 64

\bibitem[{van Eymeren {et~al.}(2009)van Eymeren, Marcelin, Koribalski, Dettmar,
  Bomans, Gach, \& Balard}]{vanEymeren:2009dh}
van Eymeren, J., Marcelin, M., Koribalski, B., et~al. 2009, A{\&}A, 493, 511

\bibitem[{Walter {et~al.}(2008)Walter, Brinks, de~Blok, Bigiel, Kennicutt,
  Thornley, \& Leroy}]{Walter:2008bq}
Walter, F., Brinks, E., de~Blok, W. J.~G., et~al. 2008, AJ, 136, 2563

\bibitem[{Werk {et~al.}(2004)Werk, Jangren, \& Salzer}]{Werk:2004de}
Werk, J.~K., Jangren, A., and Salzer, J.~J. 2004, ApJ, 617, 1004

\bibitem[{Wofford {et~al.}(2013)Wofford, Leitherer, \& Salzer}]{Wofford:2013hg}
Wofford, A., Leitherer, C., and Salzer, J. 2013, ApJ, 765, 118

\bibitem[{Zaritsky \& Rix(1997)}]{Zaritsky:1997ev}
Zaritsky, D. and Rix, H.-W. 1997, ApJ, 477, 118

\end{thebibliography}

\appendix

    \begin{table}[htbp]
   \centering
   \caption{\HI\ and stellar properties for local starburst galaxies.}
   \begin{threeparttable}
    \begin{tabular}{@{} llllllr @{}} 
        Galaxy & EW (\AA)  & L$_{\lya}$ (x10$^{41}$ erg s$^{-1}$) & fesc & M$_{HI}$(x10$^8$ \msun)  & W$_{50}$ (\kms) & M$_{\star}$(x10$^8$ \msun) \\
      \midrule
      IRAS 08339+6517     & 39.2   & 22.4  & 0.140   & 59\pom 7.4 \tnote{a}      &   201\tnote{b}  &  398.1\tnote{c}\\
      Tol 65                          & 17.5  & 0.267  & 0.053   & 7.0\pom 0.80\tnote{d} &   40\pom 6\tnote{d}    &   --\\
      NGC 6090                 & 28.3   & 6.79  & 0.027   & 160\pom 16\tnote{e}     &   211\pom 60\tnote{e}  & 2240\tnote{f}\\
      ESO 338-04               & 28.1  & 6.51  & 0.139   & 9.8\pom 1.3 \tnote{a}      &   --         &  12.6\tnote{c}\\
      LARS (Avg.)              & 21.69 &  0.91 &  0.09   & 1600                               & 227          & 177 \\ 
      \bottomrule
    \end{tabular}
        \begin{tablenotes}
        \item Note: L$_{\lya}$ and M$_{HI}$(x10$^8$) have been corrected for our cosmology. 
           \item References:  \item[a] \citet{2004ApJ...608..768C} \item[b] \citet{1991A&A...245..393M}
            \item[c] \citet{Leitet:2013cw} \item[d] \citet{Pustilnik:2007el} \item[e]\citet{vanDriel:2001gx} \item[f] \citet{Howell:2010ib}
        \end{tablenotes}
     \end{threeparttable}
    \label{tab:lyaparams}
    \end{table}

\end{document}